\documentclass{iau}

\usepackage{amsmath}
\usepackage{graphicx}
\usepackage{multirow}

\def\arxivprefixesep{:}

\newcommand{\eprint}[2][]{%
	{\tt\if!#1!#2\else#1\arxivprefixesep\ignorespaces#2\fi}%
}

\begin{document}

\lefttitle{C. Bacchini et al.}
\righttitle{The volumetric star formation law \\in nearby galaxies}

\jnlPage{1}{7}
\jnlDoiYr{2022}
\doival{xx.xxxx/xxxxx}

\aopheadtitle{Proceedings IAU Symposium 373}
\editors{Woong-Tae Kim and Tony Wong, eds.}

\title{The volumetric star formation law \\in nearby galaxies}

\author{Bacchini, C.$^{1,2,3}$; Fraternali, F.$^2$; Pezzulli, G.$^2$; Iorio, G.$^4$; Marasco, A.$^1$; Nipoti, C.$^3$}
\affiliation{
	$^1$INAF - Astronomical Observatory of Padova, Vicolo dell'Osservatorio 5, IT-35122, Padova, Italy \\email: {\tt cecilia.bacchini@inaf.it} \\
	$^2$Kapteyn Astronomical Institute, University of Groningen, Landleven 12, 9747 AD Groningen, The Netherlands \\
	$^3$Department of Physics and Astronomy, University of Bologna, via Gobetti 93/2, I-40129, Bologna, Italy \\ 
	$^4$Department of Physics and Astronomy, University of Padova, Vicolo dell’Osservatorio 3, IT-35122, Padova, Italy \\}

\begin{abstract}
Star formation laws are empirical relations between the cold gas (HI+H$_2$) content of a galaxy and its star formation rate (SFR), being crucial for any model of galaxy formation and evolution. 
A well known example of such laws is the Schmidt-Kennicutt law, which is based on the projected surface densities. 
However, it has been long unclear whether a more fundamental relation exists between the intrinsic volume densities. 
By assuming the vertical hydrostatic equilibrium, we infer radial profiles for the thickness of gaseous discs in a sample of 23 local galaxies, and use these measurements to convert the observed surface densities of the gas and the SFR into the de-projected volume densities. 
We find a tight correlation linking these quantities, that we call the volumetric star formation law. 
This relation and its properties have crucial implications for our understanding of the physics of star formation. 
\end{abstract}

\begin{keywords}
Star formation laws, ISM structure, spiral and dwarf galaxies, Milky Way.
\end{keywords}

\maketitle
\section{Introduction}
The characterisation of the relation between star formation and gas properties is crucial to understand how galaxies build their stellar mass with time. 
The most used and famous star formation is the Schmidt-Kennicutt (SK) law  \citep{1959Schmidt,2012KennicuttEvans}, which is a power-law relation with slope $N\approx 1.4$ linking the surface densities of the star formation rate (SFR, $\Sigma_\mathrm{SFR}$) and the total cold gas (HI+H$_2$, $\Sigma_\mathrm{gas}$), namely $\Sigma_\mathrm{SFR} \propto \Sigma_\mathrm{gas}^N$. 
The \textit{global} SK law involves the surface densities averaged over the whole star-forming disc, while the \textit{local} SK law is built using either azimuthally-averaged radial profiles or pixel-by-pixel measurements. 
Both the global and the local relations show a break at about $\Sigma_\mathrm{gas} \approx 10~M_\odot \mathrm{pc}^{-2}$ \citep[e.g.][]{2010Bigiel}. 
This break is typically ascribed to the presence of a density threshold for star formation, meaning that the conversion of gas into stars becomes highly inefficient in low-density regions of galaxies. 
Here, we present a new \textit{volumetric} star formation (VSF) law involving the volume densities instead of the surface densities \citep{2019Bacchini,2019Bacchini_b,2020Bacchini_b}. 
First, we describe the crucial role of the disc thickness in the derivation of the volume densities. 
Then, we show that the VSF law has no break and holds for both the high- and low-density regions of galaxies. 
Finally, we discuss the physical interpretation of our results. 

\section{The flaring of gas discs in galaxies}
Our approach considers two fundamental properties of gas discs in galaxies, which are the thickness and the flaring. 
The gas pressure is expected to balance the gravitation pull toward the galaxy midplane (vertical hydrostatic equilibrium, hereafter VHE, \citealt{1996Olling}). 
Since gravity weakens with increasing galactocentric distance, the gas layer becomes thicker moving from the inner to the outer regions of galaxies (flaring). 
This expectation is corroborated by observational works finding indications of the gas disc flaring in real galaxies \citep[e.g.][]{2011Yim,2017Marasco}. 
Unfortunately, measuring the disc thickness from observations is possible only in very favourable cases, such as the Milky Way (MW) and nearby edge-on galaxies (although often with large uncertainties).  
Thus, we calculated the gas disc thickness relying on the assumption of the VHE. 

We collected a sample of nearby galaxies including 10 spirals and 12 dwarf irregulars, and the MW. 
This sample allows us to probe not only the inner regions of spirals, which are high-density, molecular-gas rich, and high-metallicity environments, but also the spiral outskirts and the dwarf galaxies, which are low-density, atomic gas-dominated, and low-metallicity regions. 
Our goal is to convert the azimuthally-averaged radial profiles of the observed surface densities ($\Sigma_i$), which are typically used to derive the (local) SK law, into the midplane volume densities ($\rho_i$) by dividing for the scale height ($h_i$) of the disc: $\rho_i(R)=\Sigma_i(R)/(\sqrt{2\pi}h_i(R))$, where $R$ is the galactocentric distance, and $i$ is component considered (HI, H$_2$, or SFR). 
Our approach based on the VHE requires to measure the gas velocity dispersion and to calculate the gravitation potential of the galaxies. 
We accurately measured the radial profile of the gas velocity dispersion by modelling HI and CO emission line datacubes using the software $^{\text{\textsc{3D}}}$\textsc{Barolo} \cite{2015DiTeodoro}, which is based on the tilted-ring model. 	
The galactic potential is derived using a mass model made of a stellar disc and bulge, a dark matter halo, and a gaseous disc, which is fitted to the galaxy rotation curve. 
For each galaxy, we derived the potential via numerical integration using a mass model obtained though the rotation curve decomposition. 
We then used the Python tool \texttt{Galpynamics} 
\citep{2018Iorio} to calculate the scale height for both the atomic and the molecular gas discs. 
For all the galaxies, we find a significant flaring for both the gas discs. 
The last ingredient that we need is the SFR scale height, which is unfortunately not known a priori. 
Hence, as a proxy, we used a linear combination of the scale heights of the atomic gas and molecular gas discs
$h_\mathrm{SFR}(R)=h_\mathrm{HI}(R)f_\mathrm{HI}(R) + h_\mathrm{H_2}(R)f_\mathrm{H_2}(R)$, 
where the gas fractions $f_\mathrm{HI} \equiv \Sigma_\mathrm{HI}/\Sigma_\mathrm{gas}$ and $f_\mathrm{H_2}\equiv \Sigma_\mathrm{H_2}/\Sigma_\mathrm{gas}$ are used as weights. 
This prescription is based on the idea that the cold gas and young stars share the same vertical distribution. 
We compared this assumption for $h_\mathrm{SFR}$ with the scale height of classical Cepheids in the MW \citep[age$\lesssim 200$~Myr,][]{2005Bono}, finding an excellent agreement. 

\section{The volumetric star formation law}
\begin{figure}
	\centering
	\includegraphics[width=1\linewidth]{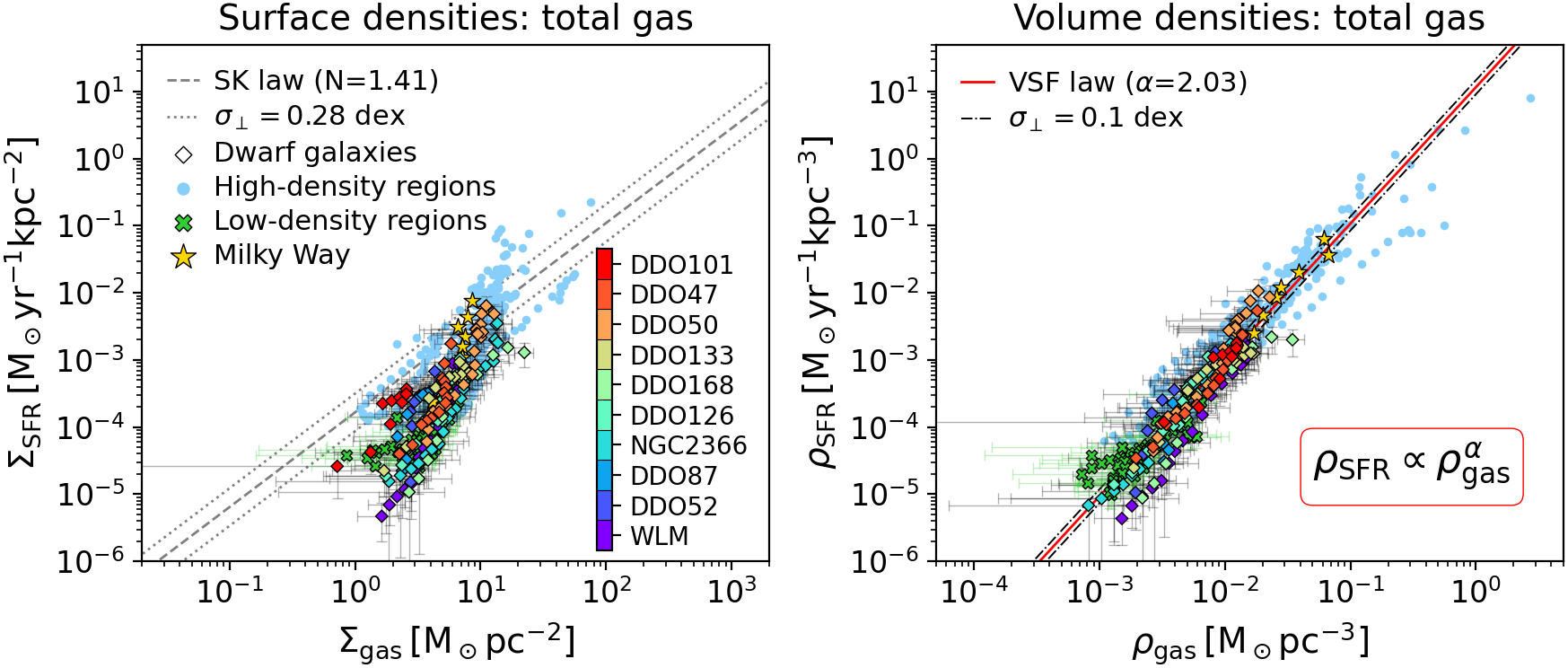}
	\caption{KS law (left) and VSF law (right) respectively based on the surface densities and the volume densities of nearby galaxies. 
		For each galaxy, the azimuthally-averaged radial profiles of the cold gas (HI+H$_2$) and SFR densities are used. 
		Diamonds are for dwarfs, which are individually tracked as indicated by the colour-bar. 
		Lightblue points and green crosses respectively show the regions within and beyond the stellar disc of spirals.
		The yellow stars are for the MW.  
		$\sigma_\perp$ is the orthogonal intrinsic scatter of the relations.}
	\label{fig:vsftesi}
\end{figure}

Having derived the volume densities for both the gas (HI and H$_2$) and the SFR, we obtained the VSF law. 
Figure~\ref{fig:vsftesi} compares the SK law and the VSF law. 
In the left panel, some galaxies tend to follow the SK law in the high-surface density regime, but most of the galaxies depart from the relation in the low-surface density regime, resulting in a large scatter ($\sigma_\perp \approx 0.3$~dex). 
The points falling out of the SK law are typically those associated with the outskirts of spirals and dwarf galaxies, where the flaring is the most prominent. 
The situation is very different when we look at the right panel in Fig.~\ref{fig:vsftesi} showing the volume densities. 
All the galaxies follow the same tight correlation, that is the VSF law. 
The best-fit relation to the volume densities is 
\begin{equation}\label{eq:log_vsf_gas}
\log \left( \frac{\rho_\mathrm{SFR}}{\mathrm{M}_\odot \mathrm{yr}^{-1} \mathrm{kpc}^{-3}} \right) = 
(2.03 \pm 0.03)
\log \left( \frac{\rho_\mathrm{gas}}{\mathrm{M}_\odot \mathrm{pc}^{-3}} \right) 
+ (1.10 \pm 0.01) \, .
\end{equation}

Three compelling points arise from these results. 
First, the VSF law has a smaller scatter ($\sigma_\perp \approx 0.1$~dex) and spans a wider dynamical range in terms of both galaxy masses and gas densities than the SK law and other surface-based relations (e.g. spatially resolved main-sequence).
These important improvements suggest that the VSF law is more fundamental than surface-based relations. 
The second point results from the absence of a break in the VSF law. 
This key difference with respect to the SK law indicates that there is no volume density threshold for star formation, at least for $\rho_\mathrm{gas} \gtrsim 10^{-3} M_\odot \mathrm{pc}^{-3}$. 
Hence, the break seen in the SK law is due to the projection effects of the disc flaring, rather than to a real drop in the star formation efficiency. 
Lastly, the slope of the VSF law disfavours the gravitational collapse as the main mechanism regulating star formation.  
Let us write the SFR volume density as $\rho_\mathrm{SFR} \propto \rho_\mathrm{gas}/\tau_\mathrm{sf}$, with $\tau_\mathrm{sf}$ being the timescale of star formation. 
The quadratic slope of the VSF law (Eq.~\ref{eq:log_vsf_gas}) implies $\tau_\mathrm{sf} \propto \rho_\mathrm{gas}^{-1}$, which is inconsistent with the gravitational free-fall time but compatible with the timescales of the gas cooling and the  H$_2$ formation, two important processes for star formation. 

To conclude, the take-home messages from this proceeding are two. 
First, the flaring of gas discs in galaxies is significant, so assuming that they are thin or have a constant thickness is not a good approximation. 
Second, the VSF law (Eq.~\ref{eq:log_vsf_gas}) is a new star formation law with smaller scatter and wider validity that the surface-based SK law, being likely more fundamental. 

\bibliographystyle{iaulike}
\bibliography{paty.bib}

\end{document}